\documentstyle[11pt,aaspp4]{article}

\begin{document}

\title{BURSTS FROM GS~1826-238: A CLOCKED THERMONUCLEAR FLASHES GENERATOR}

\author{P. Ubertini\altaffilmark{1}, A. Bazzano, M. Cocchi, and L. Natalucci}
          \affil{Istituto di Astrofisica Spaziale {\em(IAS/CNR)},
           via Fosso del Cavaliere, 00133 Roma, Italy}
\author{J. Heise, J.M. Muller\altaffilmark{2}, and J.J.M. in 't Zand}
          \affil{Space Research Organization Netherlands {\em (SRON)},
           Sorbonnelaan 2, 3584 TA Utrecht, The Netherlands}

\altaffiltext{1}{e-mail address: ubertini@alpha1.ias.rm.cnr.it}
\altaffiltext{1}{also at {\em BeppoSAX} Science Data Center, Rome, Italy}

\begin{abstract}
The transient X-ray source GS~1826-238 was monitored during five different 
observing periods between August 1996 and October 1998 with the {\em BeppoSAX} 
Wide Field Camera instrument in the framework of a deep observation of the 
Galactic Center region. A first detection of X-ray bursts from this source 
occurred, ruling out its previously suggested black hole candidacy and 
strongly suggesting the compact object to be a weakly magnetized neutron star. 
During the 2.5 years of monitoring, corresponding to $\sim 2.0$ Ms observing 
time, 70 bursts were detected from the 
source. We here report a quasi-periodicity of $5.76$~hr in the burst occurrence
time which is present during all observations. This is the first example of
quasi-periodic bursting over a period of years. It is in line with the history 
of a rather constant flux since the turn on in 1988 and points to a very stable
accretion. 

\end{abstract}


\section{INTRODUCTION}

GS~1826-238 was serendipitously detected on September 8, 1988 with an average 
X-ray  flux of about 26 mCrab in the 1-40 keV range (\cite{Maki88}). During 
September 9-16, 1988, rapid fluctuations (flickering) were observed on time 
scales down to 2 ms (\cite{Tana89}). In the 
{\em GINGA} All Sky Monitor the source was below 50 mCrab during the 
period from August through October 1988 (\cite{Tana95}). Since no detections
were available from previous X-ray observations, GS~1826-238 was 
tentatively reported as transient. Moreover, similarity with Cyg X-1 and 
GX 339-4 in the low (hard) state, both in spectrum and temporal behavior, 
suggested  to tentatively consider the source as  a black hole candidate (BHC) 
(\cite{Tana89}). The {\em GINGA} spectrum was in fact well fitted by a single 
power law with photon index $\Gamma=1.7$ ($\Gamma$ is defined through 
$N(E)=kE^{-\Gamma}$ where $N(E)$ is the number of photons per keV of 
energy $E$ and $k$ a constant).

In 1989 the source was detected by TTM on March 17 (\cite{Zand92}) at a flux
of about 32 mCrab in the energy range 2-28 keV.
Later on the source was observed in October 1990 and June and October 1992 with
the {\em ROSAT} PSPC  (\cite{Barr95}) and no bursts were detected during
8 hours of net exposure time. 
The spectrum was well fitted by a power law with a photon index in the range 
$\Gamma=1.5 \div 1.8$ plus absorption with a hydrogen column density
toward GS~1826-238 of about $N_{\rm H}=5\times10^{21}$~cm$^{-2}$.
Follow-up optical studies led to the identification of a V=19.3 optical 
counterpart located at $\alpha=18^{h}28^{m}28^{s}.2$ and $\delta=-23\arcdeg 
47\arcmin 49\farcs 12$ (equinox 2000.0). The {\em ROSAT} source was inside the 
{\em GINGA} error box as well as inside a larger one of an unidentified X-ray 
burster (\cite{VanP95}) observed with {\em OSO 7} (\cite{Mark77}) and
{\em OSO 8} (\cite{Beck76}). The error box of the unidentified X-ray
burster also contains the source 4U1831-23 (\cite{Form78}).
\cite{Barr95} point out that the faintness of the optical counterpart 
indicates that the source could be a LMXB. Given the coincidence of GS~1826-238
with the large error box of a burster and the lack of burst detections only 
weakly favours the black hole hypothesis.

During November 1994 the source was detected with OSSE at 7.5 standard 
deviations in the 60-200 keV range (\cite{Stri96}) with a spectral fit 
consistent with a power law with photon index $\Gamma=\simeq 3$. The
authors suggested
the nature of the compact object to be a neutron star (NS) in view of the 
variability characteristics of the source and a hard X-ray spectral behavior
not being unique to black hole candidates. A wide-band spectral fit had been
obtained over the 1.5-200 keV range by combining data from {\em GINGA} 
(September 1988) and OSSE (November 1994) and assuming no flux difference
between the two observations (\cite{Stri96}). The fit was well 
represented by an exponentially cut-off power law  (cut-off energy about
58 keV, photon index $\Gamma = 1.76)$ with a reflection term. The latter 
result suggested a similarity with previously studied NS systems showing at
high energies (40-200 keV) a spectral photon index of $\Gamma \sim 3$ while 
black hole candidates usually have a harder spectral photon index (around 2) 
as was also reported by \cite{Bazz96}. The suggestion that GS~1826-238 could 
contain a NS was also discussed in detail by \cite{Barr96}.\\
The recent detection of the 2.1 h optical modulation (\cite{Home98})
is a strong hint for a compact system.

We observed GS~1826-238 for 2.4~Ms during 5 periods in a time span of 
2.5~yr. This paper focuses on Primary WFC and Science Performance 
Verification data, for a total of $\sim 2.0$~Ms, showing a remarkable
quasi-periodicity in the burst occurrence times.

\section{OBSERVATIONS AND DATA ANALYSIS}

The Wide Field Cameras (WFC) (\cite{Jage97}) are designed for performing 
spatially resolved simultaneous monitoring of X-ray sources in fields of
size $40\arcdeg \times 40\arcdeg$, enabling systematic studies of spectral 
variability up to time resolutions near 1~ms. The mCrab sensitivity in 2-28 
keV over a large field of view (FOV) and the near-to-continuous operation over 
a period of years offer the unique opportunity to observe the long-term 
bursting behavior of new as well as already known (weak) sources.

This is one of the most important reasons that the Galactic Bulge is being
monitored in 1 to 2 month periods during each of the visibility periods since 
the beginning of the {\em BeppoSAX} operational life in July 1996. During 
the observations through October 1998, that combine to a total of $\sim
2.4$ Ms exposure time, at least 45 sources and about 600 bursts have been detected 
(\cite{Uber98}).

It was from the second visibility period on, between March 18 and April 
11, 1997, that the observations were analyzed in a near-to-real-time fashion
and it was then noted that GS~1826-238 exhibits X-ray bursts (\cite{Uber97}).
This is the first report of burst activity from this source.
Typical time profiles of a burst, detected on MJD 50342.54202,
is plotted in Figure 1 in three different energy 
ranges. We performed an analysis of the spectral evolution of two bursts 
detected in MJD 50340.38703 and MJD 50342.54202 through modeling 
with black body radiation; the results are given in table 1.
We show these bursts in particular because the source was close to
the center of the FOV and, thus, the sensitivity was near optimum.
The analysis points out 1) that the burst spectra are consistent with black
body radiation with temperatures between roughly 1 and 2 keV, and 2) that
there is a cooling trend during the decay of the bursts. This characterization
of the two bursts identifies them as type I X-ray bursts which are presumed
to result from a thermonuclear helium flash on the surface of a weakly
magnetized neutron star. This definitely eliminates the black hole 
hypothesis for GS~1826-238 as already suggested by Bazzano et al., 1997.

An extensive search for bursts from GS~1826-238 was performed on the whole 
data set covering August 1996 through October 1998. A total of 70 bursts were 
found. This is a large number, during the WFC observations of
the Galactic Bulge there are only two sources with more bursts (GX~354--0 and 
GRO~1744--28 that is a well known type II burster). 
A simple analysis of the burst occurrence times immediately
resulted in a surprising result. Figure 2 (top) shows a histogram of the wait times
between the 70 bursts. It has a pronounced peak at 5.76~hr and at twice that
time. None of the wait times is smaller than 5.34~hrs. Figure 2 (bottom) shows
a histogram of the wait times folded in a range of periods. A Gaussian fit 
to the distribution results in a peak at 5.76~hr and a full width at half 
maximum of 
37 min. Furthermore, the detection of $\sim 4$ burst/day during live time 
observation indicates that bursts are always present at the expected occurrence
time at the intensity level detectable by the WFC, corresponding to $\sim 100$ 
mCrab for a source at $\sim 10\arcdeg$ from the center of the FOV and a burst 
duration of $\sim 100$ s, as is almost always the case for the GS~1826-238 
observations. This is consistent with the source continuously bursting 
at times when it was monitored by WFC. 

Previous burst searches (\cite{Barr95}) carried out in order to disentangle 
the nature of this originally classified BHC (\cite{Tana89}) were unsuccessful.
This is not surprising in view of the relatively short observation times of 
these attempts, but we cannot rule out a different source state in the 
pre-{\em BeppoSAX} observations. Also, the recurrence period is almost a 
factor of 4 longer than the period of low-earth orbits that X-ray detectors
on {\em ROSAT}, {\em RXTE} and {\em BeppoSAX} are in. This increases the 
chance that whole sequences of bursts are missed by fairly long observations.

Whenever observed with the WFCs the source shows a low persistent emission of
$\sim 31$ mCrab with an intensity variation between 27 and 39 mCrab with no 
clear long term trend as shown in Figure 3. This intensity corresponds to
an average flux of 
$\sim 1.1 \times 10^{-9} {\rm erg~cm}^{-2} {\rm s}^{-1}$ in the 2-28 keV range 
(in the WFC 1 Crab corresponds to $3.6 \times 10^{-8} {\rm erg~cm}^{-2} 
{\rm s}^{-1}$ in the 2-28 keV range). The long term monitoring performed with 
the ASM on board RXTE in the same time frame as with WFC but with more uniform 
time coverage in the 2-10 keV range shows an average intensity from 1.5 to 2.9
ct/s (1ct/s is equivalent to $2.7 \times 10^{-10}  {\rm erg~cm}^{-2} 
{\rm s}^{-1}$). This is consistent with the WFC measured flux, that corresponds
to a luminosity of $\sim2.1\times10^{37}$erg~s$^{-1}$ at a distance of 
10~kpc.

\section{DISCUSSION}

A quasi-periodicity in burst occurrences is not unique to GS~1826-238: regular 
type I bursting was seen in 1658-298, 1820-303 and 1323-619 (see review by 
\cite{Lewi95} and references therein) among others. However, the stability of 
the clock for 2.5~yr is a unique feature of GS~1826-238.
There have never been hints of this in any of the other $\sim50$
known X-ray bursters
(although it should be noted that observational selection effects may play
an important role in the detection). The quasi-periodicity is an exhibition
of a stable accretion process. The fuel for the bursts must be continuously
replenished at a constant rate in order to provide the circumstances for
such a regular ignition of thermonuclear flashes. The fact that the regularity 
lasts such a long time strongly indicates that there is no reservoir of fuel 
because it is expected in general that such reservoirs will be exhausted after 
a few hundred bursts (\cite{Lamb78}) while the expected number of bursts
from GS~1826-238 during 2.5~yr for a periodicity of 5.76~hr would be of 
the order of $4\times10^3$. This notion is very useful in the
interpretation of the
so-called $\alpha$ parameter. This is the ratio of the bolometric energy of
the non-burst emission between two bursts to that of the burst following
that period. This ratio is proportional to the amount of energy per nucleon
liberated during the nuclear fusion process and can potentially constrain the
fusion process (i.e., whether hydrogen, helium or another composition of gas 
is burned).

We can determine energies of the burst and persistent emission only in the
WFC bandpass of 2 to 28 keV. As mentioned in the introduction, the persistent
emission has a hard spectrum which can only be assessed accurately in a
broad bandpass. Therefore, we extrapolate its spectrum from the WFC bandpass 
to the 1 to 58~keV band, the upper limit corresponding to the cut-off 
energy (\cite{Stri96}). The resulting average bolometric flux, corresponding 
to $31\pm4$~mCrab in 2 to 28 keV, is 
$(1.8\pm0.2)\times10^{-9}$~erg~cm$^{-2}$~erg~cm$^{-2}$ . 
In 5.76~hr this accumulates to a fluence of 
$(3.7\pm0.4)\times10^{-5}$~erg~cm$^{-2}$.
The burst energies are easier to assess because their spectra are nearly
completely contained in the WFC bandpass. We derive an average fluence of 
$(6.2\pm0.3)\times10^{-7}$~erg~cm$^{-2}$ over the two bursts here analyzed. 
Thus $\alpha=60\pm7$. There are a few effects which can bias $\alpha$,
see Lewin, Van Paradijs \& Taam (1993).. Through the quasi-periodicity one
can at least eliminate one of them: that a reservoir of fuel would suppress
$\alpha$. The value of $\alpha$=60 is commonly observed and confirms the
picture that helium is burned during the flash.

In table 2, a history of GS~1826-238 fluxes since the 1988 discovery is 
presented. This indicates that since 1988 the source behaves as a weak, 
persistent source showing variability of a few tens of percents at
most in the 1 to 30 or 40 keV energy band. Given the low flux and the meager 
coverage of the sky position by sensitive X-ray telescopes prior to 1988, one 
may be prompted to question the transient nature of GS~1826-238. Prior to 1988 
the position was covered by a number of survey instruments such as on
{\em UHURU} (\cite{Form78}) and {\em ARIEL V} (\cite{Warw81}). These 
instruments surveyed the sky during the 1970s. The source 
detected nearest to GS~1826-238 was 4U~1831-23, at an angular separation of
1\fdg4. Given the small error box of 4U~1831-23, this implies that both 
sources cannot be the same and that GS~1826-238 was not detected (see also
\cite{Barr95}). From the flux numbers for 4U~1831-23 we estimate conservatively
the upper limit for GS~1826-238 to be $\sim$2~mCrab in 2 to 10 keV, that is one
order of magnitude smaller than after 1988. We conclude that, since then, 
GS 1826-238 is a weak (20 to 40 mCrab) persistent source with a very stable
accretion flow of $\sim 1.5 \times10^{-9}{\rm M}_{\sun}$/y, at 10 kpc.\\   

We are continuing the analysis of the whole data set, in particular to 
characterize each individual burst with respect to burst duration, fluence, 
and peak flux and study the correlation between the burst properties and the 
level of persistent emission and the wait time. Furthermore, an analysis 
of a sensitive observation with the narrow-field instruments on BeppoSAX in 
ongoing (In 't Zand et al., in preparation).

Acknowledgments

\acknowledgments
The authors thank Dr. D. Barret and Prof. J. van Paradijs for very useful
scientific discussions and suggestions.

We also thank Team Members of the {\em BeppoSAX} Science Operation Centre
and Science Data Centre for their continuous support and timely actions for
quasi-$"$real time$"$ detection of new transient and bursting sources and
the follow-up TOO observations.



\begin{deluxetable}{crrrrrrrrrrr}

\footnotesize
\tablecolumns{8}
\tablewidth{0pt}
\tablecaption{Burst parameters} \label{tbl2}
\tablehead{
\colhead{}    &  \multicolumn{3}{c}{MJD 50340 burst} &   \colhead{}   &
\multicolumn{3}{c}{MJD 50342 burst} \\
\cline{2-4} \cline{6-8} \\
\colhead{Burst Period} & \colhead{Total}   & \colhead{Peak}    & \colhead{Tail} &  \colhead{} &
                         \colhead{Total}   & \colhead{Peak}    & \colhead{Tail}  }
\startdata
Integration Time (s) & $56$ & $13$ & $43$  &&  $56$ & $13$ & $43$\\
Temperature (keV)    & $2.05 \pm 0.07$ & $2.26 \pm 0.09$ & $1.92 \pm 0.08$  &&
                       $1.81 \pm 0.06$ & $2.10 \pm 0.08$ & $1.59 \pm 0.09$  \\
$R_{\rm bb}$ (km) at 10 kpc&$6.8\pm0.5$& $7.1\pm0.6$   & $7.0\pm0.6$     &&
                       $8.5\pm0.6$     & $8.3\pm0.7$   & $9.7\pm1.1$ \\
$\chi^{2}_{{\rm d.o.f.}}$ (27 d.o.f.) & $0.77$ & $0.89$ & $0.87$  &&
                                        $1.22$ & $0.99$ & $1.59$ \\

\cutinhead {Burst {\em e}-folding time (s) in different energy bands}


$2-28$ keV & & $35.9 \pm 4.2 $   & &   &&  $41.8\pm4.5$   & \\
$2-7$  keV  &  & $52.6 \pm 8.8$ &  &   &&
                       $53.7\pm 7.2$   &   \\
$7-28$ keV & & $22.7\pm3.5$&    &      &&
                       $22.0\pm3.7$         &  \\

\enddata

\end{deluxetable}

\begin{deluxetable}{crrrrrrrrrrr}
\footnotesize
\tablecaption{History of the GS~1826-238 observed fluxes} \label{tbl2}
\tablewidth{0pt}
\tablehead{
\colhead{Energy range (keV)} & \colhead{Instrument/Satellite}  &
\colhead{Year}  & \colhead{Flux (erg ${\rm cm}^{-2} {\rm s}^{-1}$)} &
\colhead{Reference}
}
\startdata
$2-27$ & ASM/{\em GINGA} & $1988$ & $1.26 \times 10^{-9}$ & \cite{Tana95}  \\
$1-40$ & LAPC/{\em GINGA} & $1988$ & $1.06 \times 10^{-9}$ & \cite{Tana89}  \\
$1-40$ & TTM/{\em MIR} & $1989$ & $1.1 \times 10^{-9}$ & \cite{Zand92}  \\
$0.1-2.4$ & {\em ROSAT} & $1990,1992$ & $1.24 \times 10^{-10}$ & \cite{Barr95}  \\
$2-10$ & ASM/{\em RXTE} & $1996-1998$ & $0.41\div 0.77 \times 10^{-9}$ & {\rm Public Data}   \\
$2-28$ & WFC/{\em BeppoSAX} & $1996-1998$ & $1.01 \div 1.46 \times 10^{-9}$ & \cite{Uber98}  \\
\enddata

\tablenotetext{a}{The {\em ROSAT} value is in the soft band and the flux is not directly
  comparable with the other measurements}

\end{deluxetable}

\clearpage

\begin{figure}
 \plotone{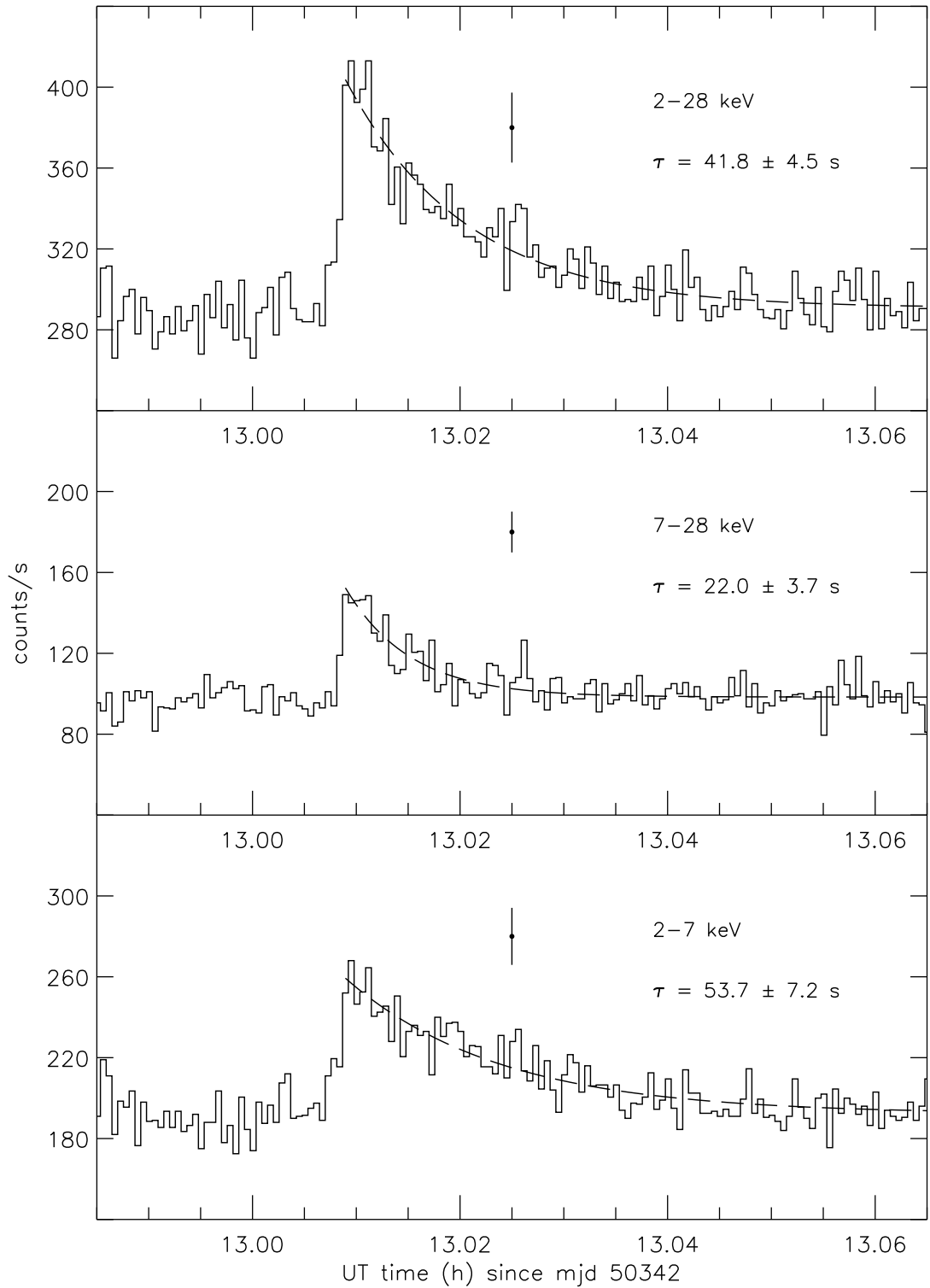}
\caption{
     The MJD 50342.54202 burst in three energy bands. The dashed line is the exponential decay fit.
    \label{fig1}}
\end{figure}

\begin{figure}
 \plotone{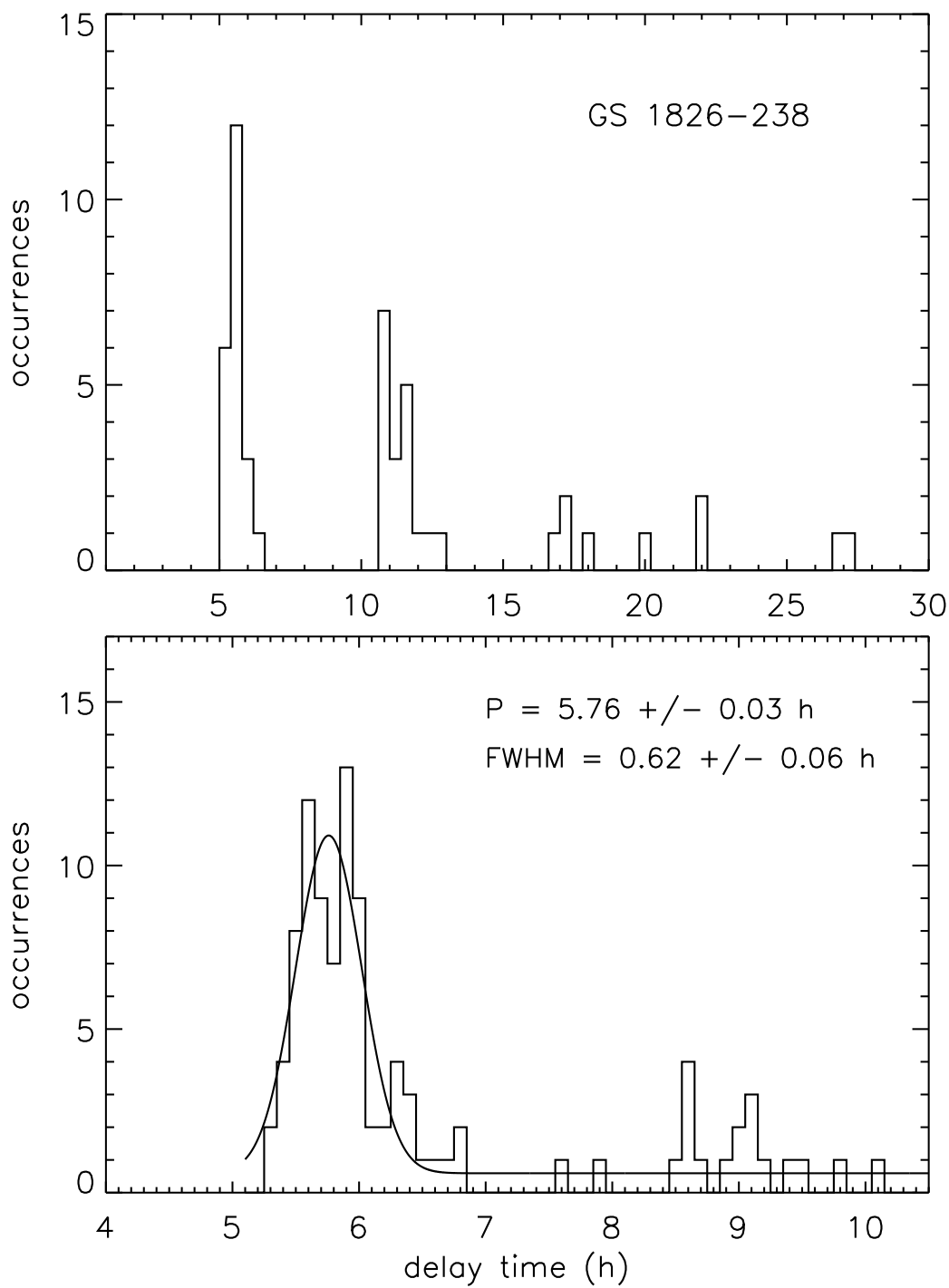}
\caption{
      Top: number of GS~1826-238 burst occurrences as a function of the delay 
      time since the last observed burst. Bottom: a histogram obtained by 
      folding the observed delay times to periods ranging from 5.3 
      (minimum observed delay) to 10.5 hr 
      with 0.1 hr time resolution. The best gaussian fit, whose parameters 
      are shown in the figure, is also plotted.
    \label{fig2}}
\end{figure}

\begin{figure}
 \plotone{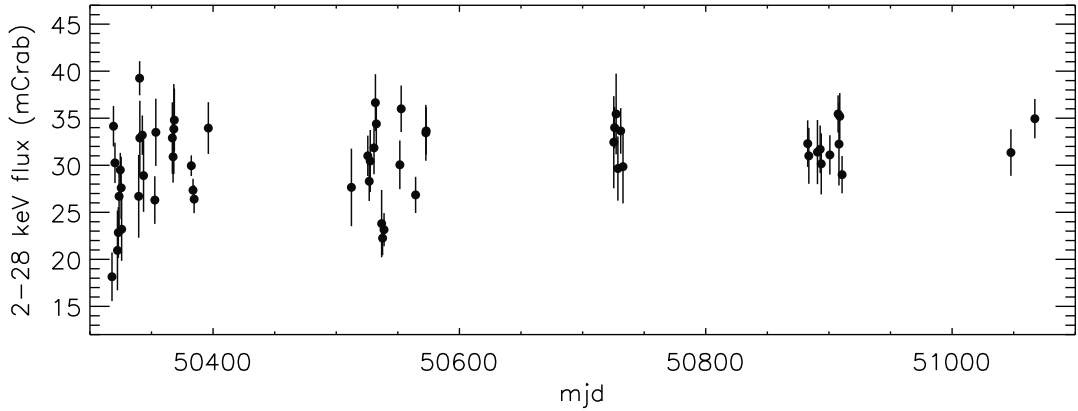}
\caption{
     Time history of the persistent emission of GS~1826-238 in the WFC
     energy range
     (2-28 keV). The weight average flux is 31 mCrab and corresponds to  
     $1.4\times10^{37}\rm{erg~s}^{-1}$ (at d=10 kpc, $N_{\rm H}=5\times10^{21}$~cm$^{-2}$ and 
     $\Gamma=1.76$). 
The luminosity becomes $2.1\times10^{37}\rm{erg~s}^{-1}$ when
integrating the spectrum in the range 2-58 keV, the upper value
corresponding to the observed high energy cut-off (Strickman et al.,
1996).
    \label{fig3}}
\end{figure}


\clearpage

\begin{thebibliography}{}

\bibitem[Barret et al. 1995]{Barr95}
         Barret, D., Motch, C., and Pietch, W., 1995, \aap, 303, 526
\bibitem[Barret et al. 1996]{Barr96}
         Barret, D., Mc Clintock, J.E., and Grindlay, J.E., 1996, \apj, 473, 963
\bibitem[Bazzano et al.. 1996]{Bazz96}
         Bazzano A. et al., 1996,  ESA SP 382, 261, "The Transparent Universe, 2nd
         Integral Workshop"
\bibitem[Bazzano et al. 1997]{Bazz97}
         Bazzano A. et al., 1997,  AIP Conf. Proc. N. 410, 729, 1997
\bibitem[Becker et al. 1976]{Beck76}
         Becker R.H., et al. 1976, IAU circ. 2953
\bibitem[Forman et al. 1978]{Form78}
         Forman W. et al., 1978, Ap. J. Suppl., 38, 357
\bibitem[Homer et al 1998]{Home98}
         Homer L., Charles, P.A. and O'Donogue, D., Mon. Not. Astron. Soc.,  1998 (in press)
\bibitem[Jager et al. 1997]{Jage97}
         Jager, R., et al. 1997, \aap, 125, 557
\bibitem[in 't Zand 1992]{Zand92}
         in 't Zand J.J.M., 1992, Ph.D. thesis, Utrecht University
\bibitem[in 't Zand 1999]{Zand99}
         in 't Zand J.J.M., 1999, in preparation
\bibitem[Lamb \& Lamb 1978]{Lamb78}
         Lamb D.Q. and Lamb, F.K., 1978, Ap. J., 220, 291
\bibitem[Lewin, van Paradijs, \& Taam 1995]{Lewi95}
         Lewin, W.H.G., van Paradijs, J., \& Taam, R.E. 1995, in {\em "X-ray Binaries"},
         ed. W. Lewin, J. van Paradijs, \& E. van den Heuvel,
         Cambridge University Press, Cambridge, p. 175
\bibitem[Makino et al. 1988]{Maki88}
         Makino F., et al. 1988, IAUC 4653.
\bibitem[Markert et al. 1977]{Mark77}
         Markert T. H., et al., 1977,  Ap.J. 218, 801
         Ruderman M., 1972, Ann. Rev. Astr. Ap., 10, 427
       Shapiro, S.L. and Teukolsky S.A., 1983, Black Holes, White Dwarfs and Neutron Stars (New York:Wiley), 348
         Smale, A.P., Zhang, W., and White, N.E., 1997, \apj, 483, L119
\bibitem[Strickman et al. 1996]{Stri96}
         Strickman M. et al., 1996, \aap Suppl. Ser., 120, 217
\bibitem[Tanaka 1989]{Tana89}
         Tanaka Y., 1989, proceeding  of 23rd ESLAB Symp, pag. 3
\bibitem[Tanaka \& Lewin 1995]{Tana95}
         Tanaka Y. \& Lewin, W.G.A., 1995, in X-ray Binaries, Cambridge University Press, Series
         26, pag. 126
\bibitem[Ubertini et al. 1997]{Uber97}
         Ubertini P., et al., 1997, IAU circ. 6611
\bibitem[Ubertini et al. 1998]{Uber98}
         Ubertini P.,  et al., 1998, Proceedings of "The Extreme Universe, 3rd Integral
         Workshop", Taormina, Italy, 14-18 Sept. 1998, in press
\bibitem[Warwick et al. 1981]{Warw81}
         Warwick R.S.  et al., 1981, MNRAS 197, 865
         van Paradijs, J., and Lewin, W.H.G., 1985, \aap, 157, L10
\bibitem[van Paradijs 1995]{VanP95}
         van Paradijs J., 1995, in {\em "X-Ray Binaries"} ed. W. Lewin, J. van Paradijs \& E. van den Heuvel,
         Cambridge University Press, Cambridge, p. 536
         Zhang W., Stromayer, T.E. and Swank, J.H., 1997, \apj, 482, L167

\end{thebibliography}
\end{document}